\documentclass[apj]{emulateapj}

\usepackage{amsmath, amssymb}
\usepackage{mathptmx}
\usepackage{verbatim}
\usepackage{graphicx}
\usepackage{epstopdf}
\usepackage{textcomp}
\usepackage{latexsym}
\usepackage{multirow}
\usepackage{wasysym}
\usepackage[backref,breaklinks,colorlinks,citecolor=blue]{hyperref}
\usepackage{fixltx2e}


\slugcomment{Received 2019 October 31; Accepted 2020 February 17}

\shorttitle{Prescriptions for Correcting UV-Based Redshifts}
\shortauthors{Dix et al.}

\begin{document}

\title{Prescriptions for Correcting Ultraviolet-Based Redshifts for Luminous Quasars at High Redshift}

\author{Cooper Dix,\altaffilmark{1}
Ohad Shemmer,\altaffilmark{1}
Michael S. Brotherton,\altaffilmark{2}
Richard F. Green,\altaffilmark{3}
Michelle Mason,\altaffilmark{2}
Adam D. Myers,\altaffilmark{2}
}

\altaffiltext{1}
		   {Department of Physics, University of North Texas, Denton, TX 76203, USA; cooperdix@my.unt.edu}
\altaffiltext{2}
		   {Department of Physics and Astronomy, University of Wyoming, Laramie, WY, 82071, USA}
\altaffiltext{3}
                     {Department of Astronomy, University of Arizona, Tucson, AZ, 85721, USA}
                     
\begin{abstract}
~~ High-redshift quasars typically have their redshift determined from rest-frame ultraviolet (UV) emission lines.  However, these lines, and more specifically the prominent C~{\sc iv} $\lambda 1549$ emission line, are typically blueshifted yielding highly uncertain redshift estimates compared to redshifts determined from rest-frame optical emission lines.  We present near-infrared spectroscopy of 18 luminous quasars at \hbox{$2.15 < z < 3.70$} that allows us to obtain reliable systemic redshifts for these sources.  Together with near-infrared spectroscopy of an archival sample of 44 quasars with comparable luminosities and redshifts, we provide prescriptions for correcting UV-based redshifts.  Our prescriptions reduce velocity offsets with respect to the systemic redshifts by $\sim140$ km s$^{-1}$ and reduce the uncertainty on the UV-based redshift by $\sim25\%$ with respect to the best method currently used for determining such values.  We also find that the redshifts determined from the Sloan Digital Sky Survey Pipeline for our sources suffer from significant uncertainties, which cannot be easily mitigated.  We discuss the potential of our prescriptions to improve UV-based redshift corrections given a much larger sample of high redshift quasars with near-infrared spectra.
\end{abstract}
\keywords{galaxies: active --- quasars: emission lines --- quasars}

\section{Introduction} \label{sec:intro}

The best practical indicators for a quasar's systemic redshift ($z_{\rm sys}$) lie in the rest-frame optical band, particularly the prominent [O~{\sc iii}] $\lambda 5007$, Mg~{\sc ii} $\lambda 2800$, and the Balmer emission lines \citep[e.g.,][]{2005AJ....130..381B,2016ApJ...831....7S}.  However, at high-redshift ($z \gtrsim0.8$), \hbox{$\approx10^{5}$} quasars typically have their $z_{\rm sys}$ values determined from rest-frame ultraviolet (UV) spectra since only $0.1\%$ of these quasars have corresponding rest-frame optical information from near-infrared (NIR) spectra \citep[e.g.,][]{2010AJ....139.2360S,2017A&A...597A..79P,2018A&A...613A..51P}.  Unfortunately, the UV-based $z_{\rm sys}$ estimates are highly inaccurate and imprecise given that the UV emission lines are usually blueshifted by up to $\approx3000$ km s$^{-1}$ \citep[e.g.,][]{1982ApJ...263...79G,1992ApJS...79....1T,2009ApJ...692..758G,2016ApJ...831....7S}.  Mitigating these biases requires identifying robust corrections to UV-based redshifts.\\
\indent Reliable redshift estimates are needed for multiple reasons.  For example, accurate redshift estimates provide information on the kinematics of the outflowing material in the vicinity of the supermassive black hole, which likely impacts the star formation rate in the quasar's host galaxy \citep[e.g.,][]{2010MNRAS.401....7H}.  Additionally, various cosmological studies utilize conversions between redshift differences and distances \citep[e.g.,][]{1999astro.ph..5116H,2019MNRAS.482.3497Z}.  In this context, a velocity offset of 500 km s$^{-1}$ corresponds to a comoving distance of $\approx5h^{-1}$ Mpc at $z=2.5$, which can impact our understanding of, e.g., quasar clustering as velocity offsets can be misinterpreted to be distances in the redshift direction \citep[e.g.,][]{2013JCAP...05..018F,2013ApJ...776..136P}.\\
\indent The Sloan Digital Sky Survey \citep[SDSS;][]{2000AJ....120.1579Y} provides observed-frame optical spectra and redshifts forhundreds of thousands of quasars.  The redshifts determined for these quasars stem from a cross-correlation by a composite quasar template spectrum provided by \citet{2001AJ....122..549V}.  However, these estimates become increasingly uncertain in high-redshift quasars because mostly rest-frame UV emission lines are present in the optical band.  The first meaningful correction to these UV-based redshifts was achieved by \citet[][hereafter HW10]{2010MNRAS.405.2302H}.  They achieved this by introducing a two-part linear relation between the absolute magnitude and redshift of quasars.  A more recent improvement to the HW10 method was achieved by \citet[][hereafter M17]{2017MNRAS.469.4675M}, by comparing [O~{\sc iii}]-based $z_{\rm sys}$ values with the spectral properties of the C~{\sc iv} $\lambda 1549$ emission line for 45 quasars with $z \gtrsim 2.2$.\\
\indent In this work, we expand on the M17 method by adding high quality NIR spectra of 18 quasars at \hbox{$2.15 < z < 3.70$.}  We perform multiple regression analyses and provide improved prescriptions for correcting a variety of UV-based redshifts when the C~{\sc iv} line is available in the spectrum.
\indent This paper is organized as follows.  In Section \ref{sec:2}, we describe our sample selection, observations, and data analysis.  In Section \ref{sec:3}, we present our spectroscopic measurements and in Section \ref{sec:4} we discuss our results.  Our conclusions are presented in Section \ref{sec:5}.  Throughout this paper, we compute luminosity distances using $H_{0} = 70$ km s$^{-1}$ Mpc$^{-1}$, $\Omega_{\rm M} = 0.3$, and $\Omega _{\Lambda} = 0.7$ \citep[e.g.,][]{2007ApJS..170..377S}.

\begin{deluxetable*}{lccccclc}
\tablecaption{Observation Log \label{tab:log}}
\tablehead
{
\colhead{} & \colhead{}  & \colhead{}&
\colhead{} & \colhead{$H$\tablenotemark{c}} & \colhead{$K$\tablenotemark{c}} & \colhead{} & \colhead{Net Exp.} \\
\colhead{Quasar} & \colhead{$z$}  & \colhead{$z_{\rm ref}\tablenotemark{a}$}&
\colhead{$z_{\rm sys}\tablenotemark{b}$} & \colhead{(mag)} & \colhead{(mag)} & \colhead{Obs. Date} & \colhead{(s)}
}
\startdata
 SDSS J013435.67$-$093102.9 & 2.225 & 1 & 2.214 & 14.8 & 13.6 & 2016 Aug 25 & 2880\\
 SDSS J014850.64$-$090712.8 & 3.303 & 1 & 3.329 & 16.7 & 15.5 & 2016 Sep 19 & 4800\\
 SDSS J073607.63$+$220758.9\tablenotemark{d} & 3.464 & 2 & 3.445 & 16.1 & 14.9 & 2016 Sep 20 & 3840\\
 \nodata & \nodata & \nodata &\nodata & \nodata & \nodata & 2016 Sep 22 & 3840\\
 SDSS J142243.02$+$441721.2 & 3.530 & 1 & 3.651\tablenotemark{e} & 15.2 & 14.4 & 2016 Sep 7 & 1920\\
 SDSS J153750.10$+$201035.7 & 3.413 & 3 & 3.413 & 15.7 & 15.4 & 2016 Sep 22 & 3840\\
 SDSS J153830.55$+$085517.0 & 3.563 & 1 & 3.550 & 15.6 & 14.6 & 2016 Sep 19 & 1920\\
 SDSS J154359.43$+$535903.1\tablenotemark{d} & 2.379 & 1 & 2.364 & 15.0 & 14.2 & 2016 Sep 21 & 2880\\
 SDSS J154446.33$+$412035.7\tablenotemark{d} & 3.551 & 1 & 3.567\tablenotemark{e}  & 15.6 & 15.5 & 2016 Sep 20 & 3840\\
 SDSS J154938.71$+$124509.1 & 2.377 & 4 & 2.369 & 14.5 & 13.5 & 2016 Sep 5 & 1920\\
 SDSS J155013.64$+$200154.5 & 2.196 & 1 & 2.188 & 15.1 & 14.2 & 2016 Sep 19 & 2400\\
 SDSS J160222.72$+$084538.4\tablenotemark{d} & 2.276 & 1 & 2.275 & 15.0 & 14.0 & 2016 Sep 6 & 2880\\
 SDSS J163300.13$+$362904.8\tablenotemark{d} & 3.575 & 1 & 3.570 & 15.5 & 15.1 & 2016 Sep 22 & 2640\\
 SDSS J165137.52$+$400218.9 & 2.342 & 1 & 2.338 & 15.0 & 13.7 & 2016 Sep 6 & 2880\\
 SDSS J172237.85$+$385951.8 & 3.390 & 2 & 3.367 & 16.0 & 15.3 & 2016 Sep 19 & 3840\\
 SDSS J210524.47$+$000407.3\tablenotemark{d} & 2.307 & 1 & 2.344\tablenotemark{e} & 14.7 & 13.8 & 2016 Aug 26 & 1920\\
 SDSS J212329.46$-$005052.9 & 2.268 & 1 & 2.270\tablenotemark{f} & 14.6 & 13.9 & 2016 Sep 5 & 1920\\
 SDSS J221506.02$+$151208.5 & 3.285 & 2 & 3.284 & 16.4 & 15.2 & 2016 Aug 26 & 3840\\
 SDSS J235808.54$+$012507.2 & 3.401 & 2 & 3.389 & 14.7 & 13.8 & 2016 Aug 26 & 2880
\enddata
\tablenotetext{a}{(1) HW10; (2) \citet{2014ApJS..215...12C}; (3) \citet{2009ApJS..180...67R}; (4) \citet{2006AJ....131..680H}.}
\tablenotetext{b}{Unless otherwise noted, the systemic redshift was obtained from the peak of the [O~{\sc iii}] $\lambda 5007$ emission line, where available, as explained in the text.  Uncertainties on these values, discussed in Section \ref{sec:Fit}, average $\sim150$ km s$^{-1}$.}
\tablenotetext{c}{Vega-based magnitudes were obtained from 2MASS.}
\tablenotetext{d}{Indicates a BAL quasar.}
\tablenotetext{e}{Systemic redshift was determined from $\lambda_{\rm peak}$ of the H$\beta$ emission line.}
\tablenotetext{f}{Systemic redshift was determined from $\lambda_{\rm peak}$ of the Mg~{\sc ii} emission line from the SDSS spectrum of the source.}
\end{deluxetable*}

\begin{figure*}
\plotone{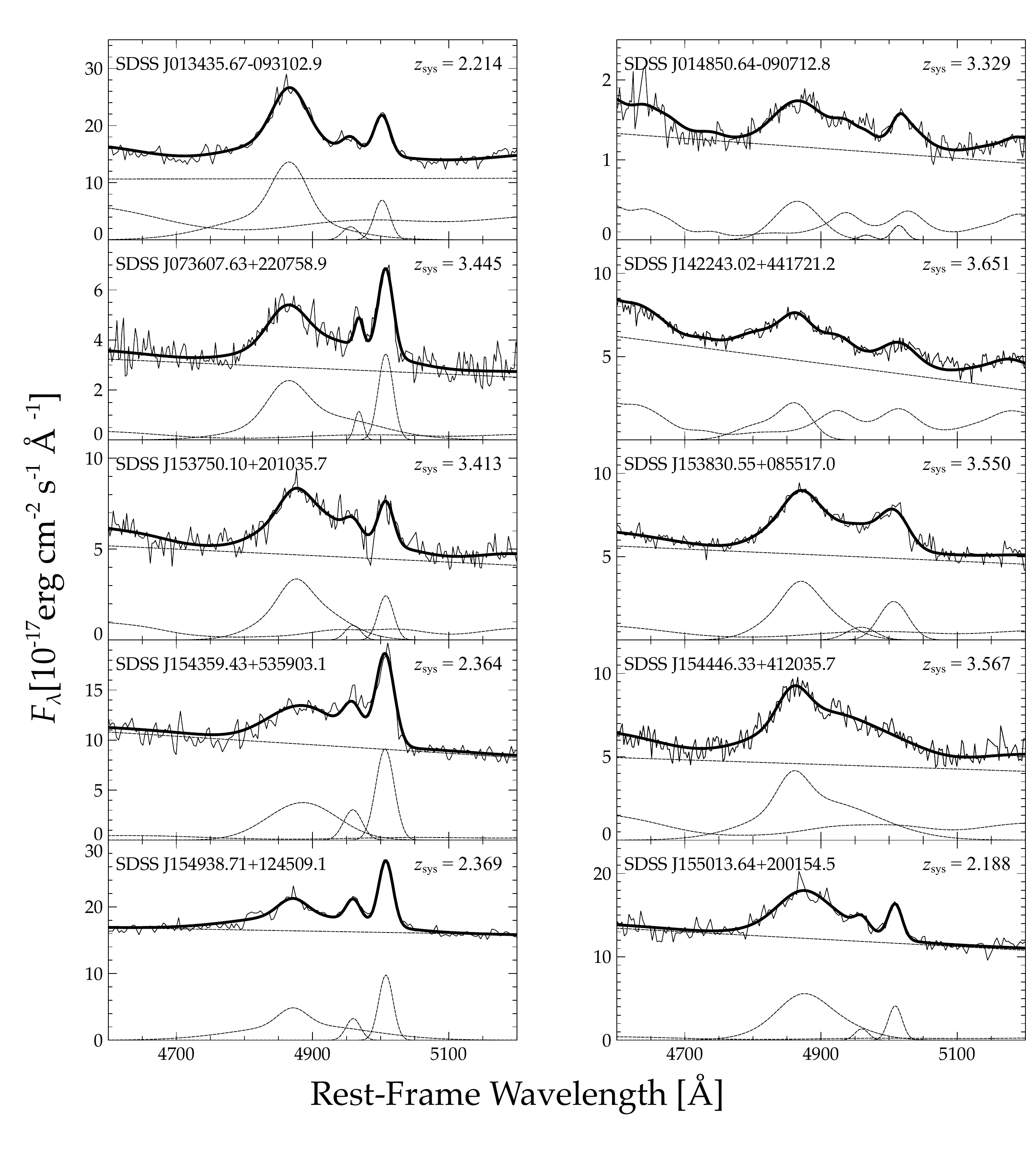}
\caption{NIR spectra of $2.15 < z < 3.70$ quasars.  The spectrum in each panel is given by a thin solid line.  The fit to each individual feature, Fe~{\sc ii}, H$\beta$, and [O~{\sc iii}] where applicable, and the linear continuum are indicated by dashed lines.  The overall fit to each spectrum is given by the bold solid line. \label{fig:fit1}}
\end{figure*}[h]
\begin{figure*}
\figurenum{1}
\plotone{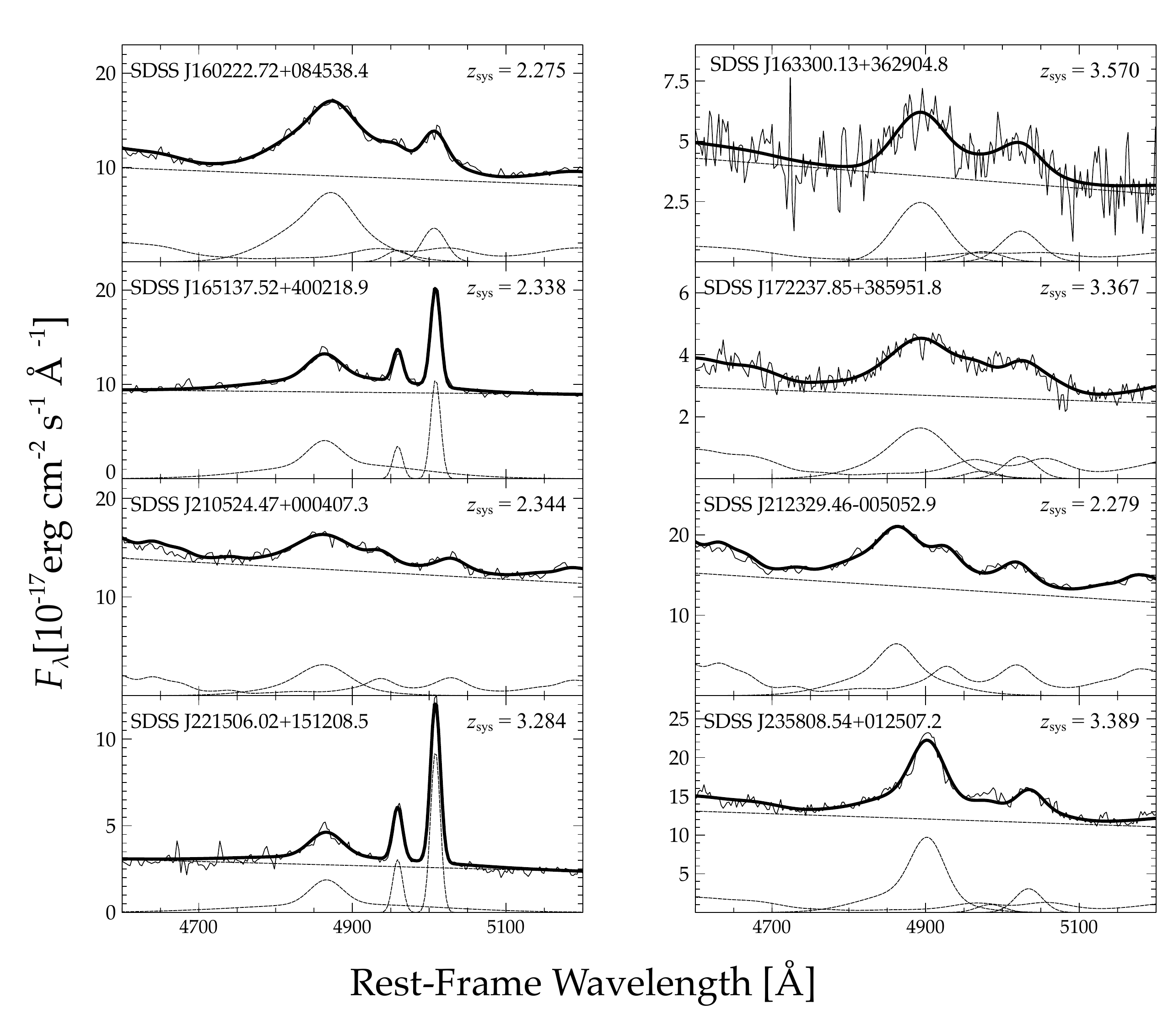}
\caption{(Continued.)}
\end{figure*}
\section{Sample Selection, Observations, and Data Analysis} \label{sec:2}

We have selected a sample of 18 quasars for our investigation based upon the following criteria:\\

\begin{enumerate}
\item Availability of a flux-calibrated optical spectrum from the SDSS recorded in the Data Release~10 quasar catalog \citep{2014A&A...563A..54P}.
\item Brightness in the range $m_{i} < 18.5$ in order to keep the signal-to-noise (S/N) ratio of the H$\beta$ region of the respective NIR spectrum, obtained with a 3.8\,m telescope, at $\approx40$.
\item Redshift within one of the following intervals, $2.15~<~z~<~2.65$\footnote{This redshift interval ensures spectral converage also of the H$\alpha$ emission line in the $K$ band.} and $3.20 < z < 3.70$, in which, at a minimum, the H$\beta$ and [O~{\sc iii}] lines can be modeled accurately within one of the near-infrared transmission windows in the $H$ or $K$ bands.
\end{enumerate}
\begin{deluxetable*}{lccccccccc}
\tablecaption{Spectral Measurements of the H$\beta$ Region and H$\alpha$ \label{tab:beta}}
\tablewidth{0pt}
\tablehead{
\colhead{} & \colhead{FWHM\textsubscript{H$\beta$}} & \colhead{EW\textsubscript{H$\beta$}} & \colhead{$\lambda$\textsubscript{peak~H$\beta$}}
& \colhead{FWHM\textsubscript{[O~{\sc iii}]}} &\colhead{EW\textsubscript{[O~{\sc iii}]}} & \colhead{$\lambda$\textsubscript{peak~[O~{\sc iii}]}\tablenotemark{a}} & \colhead{FWHM\textsubscript{H$\alpha$}} & \colhead{EW\textsubscript{H$\alpha$}} & \colhead{$\lambda$\textsubscript{peak~H$\alpha$}} \\
\colhead{Quasar} & \colhead{(km s$^{-1}$)}  & \colhead{(\AA)} & \colhead{{(\AA)}} & \colhead{(km s$^{-1}$)}  & \colhead{(\AA)} & \colhead{(\AA)}  & \colhead{(km s$^{-1}$)}  & \colhead{(\AA)} & \colhead{{(\AA)}}
}
\startdata
 SDSS J013435.67$-$093102.9 & 4438 & 99.7 & 15656 & 1625 & 14.6 & 16091 
  & 2882 & 444 & 21125 \\
 SDSS J014850.64$-$090712.8 & 4716 & 33.7 & 21035 &
 1513 & 4.3 & 21680 & \nodata & \nodata & \nodata\\
 SDSS J073607.63$+$220758.9\tablenotemark{b}  & 6876 & 94.3 & 21625 &
 1640 & 31.6 & 22256 & \nodata & \nodata & \nodata\\
 SDSS J142243.02$+$441721.2 & 4563 & 39.9 & 22607 &
 \nodata & \nodata & \nodata & \nodata & \nodata & \nodata\\
 SDSS J153750.10$+$201035.7 & 5107 & 69.5 & 21516 &
 1613 & 14.6 & 22094 & \nodata & \nodata & \nodata\\
 SDSS J153830.55$+$085517.0 & 5512 & 70.8 & 22161 &
  3192 & 26.1 & 22782  & \nodata & \nodata & \nodata \\
 SDSS J154359.43$+$535903.1 & 8301 & 54.3 & 16495 & 
 1835 & 28.6 & 16843 & 7495 & 543 & 22171 \\
 SDSS J154446.33$+$412035.7 & 7235 & 132.4 & 22202 &
 \nodata & \nodata & \nodata & \nodata & \nodata & \nodata \\
 SDSS J154938.71$+$124509.1 & 5495 & 42.4 & 16408 &
 1544 & 15.4 & 16866 & 5550 & 374 & 22139\\
 SDSS J155013.64$+$200154.5 & 6539 & 61.9 & 15544 &
 1325 & 7.5 & 15960 & 5178 & 391 & 20962\\
 SDSS J160222.72$+$084538.4 & 6676 & 122.3 & 15951 &
 2387 & 19.5 & 16398 & 5629 & 586 & 21517\\
 SDSS J163300.13$+$362904.8 & 4876 & 57.8 & 22297 &
 3768 & 24.6 & 22884& \nodata & \nodata & \nodata \\
 SDSS J165137.52$+$400218.9 & 4405 & 65.6 & 16234 &
 957.8 & 18.5 & 16713 & 4380 & 377 & 21920 \\
 SDSS J172237.85$+$385951.8 & 5938 & 67.9 & 21300 &
 3028 & 13.9 & 21866 & \nodata & \nodata & \nodata \\
 SDSS J210524.47$+$000407.3 & 5331 & 25.3 & 16256 & \nodata & \nodata & \nodata & 4530 & 281 & 21975\\
 SDSS J212329.46$-$005052.9 & 4500 & 48.1 & 15929 & \nodata & \nodata & \nodata & 4084 & 319 & 21540\\
 SDSS J221506.02$+$151208.5 & 4059 & 100.0 & 20840 & 
 956.9 & 61.7 & 21450 & \nodata & \nodata & \nodata\\
 SDSS J235808.54$+$012507.2 & 3702 & 63.3 & 21397 &
 2652 & 11.6 & 21974& \nodata & \nodata & \nodata 
\enddata
\tablenotetext{a}{Corresponding to the [O~{\sc iii}]~$\lambda 5007$ component.}
\tablenotetext{b}{SDSS J073607.63$+$220758.9 was observed on two different nights, as denoted in Table \ref{tab:log}, and, therefore, we present the values stemming from the stacked spectrum.}
\end{deluxetable*}
Spectroscopic observations of this sample were performed at the United Kingdom Infrared Telescope (UKIRT) on Mauna Kea, Hawaii.  The observation log and quasar basic properties appear in Table \ref{tab:log}.\\
\indent We utilized the UKIRT Imager-Spectrometer (UIST) with a slit width of 0.24\arcsec\ to maximize the resolution at the expense of potentially higher slit losses.  During these observations, the telescope was nodded in an ABBA pattern in order to obtain primary background subtraction.  The broad band B2 filter was used in order to obtain a wavelength range of approximately $1.395 - 2.506~ \mu$m, spanning the $H$ and $K$ bands as necessary.  The dispersion for these observations was $10.9$ {\AA} pixel$^{-1}$ with a spectral resolution of $R \sim 448$.  Standard stars of spectral type G and F were observed on each night alongside the quasar in order to remove the telluric features that are present in the quasars' spectra.\\
\indent The two-dimensional spectra of the quasars and the standard stars were obtained using standard IRAF\footnote{IRAF (Image Reduction and Analysis Facility) is distributed by the National Optical Astronomy Observatory, which is operated by AURA, Inc., under cooperative agreement with the National Science Foundation.} routines.  Each of the objects was initially pair subtracted in order to remove most of the background noise.  Then, both the positive and negative residual peaks were analyzed and averaged together.  During the analysis, wavelength calibration was achieved using Argon arc lamps.  The hydrogen features in each standard star were removed prior to removing the telluric features from the quasars' spectra.\\
\indent Removal of the telluric features and the instrumental response from the quasar spectra was done by dividing these spectra by their respective standard star spectra.  Then, any remaining cosmic ray signatures on the quasar spectra were carefully removed.  Final, flux calibrated quasar spectra were obtained by multiplying these data by blackbody curves with temperatures corresponding to the spectral types of the telluric standards and by a constant factor that was determined by comparing the $H$, for $2.15 < z < 2.65$, or $K$, for $3.20 < z < 3.70$, band magnitudes from the Two Micron All Sky Survey \citep[2MASS;][]{2006AJ....131.1163S} to the integrated flux across the respective band using the flux conversion factors from Table A.2 of \citet{1998A&A...333..231B}.  We do not rely on the telluric standards for the purpose of flux calibration given the relatively narrow slit and the differences in atmospheric conditions between the observations of the quasars and their respective standard stars.  For each source, we utilized their SDSS spectrum to verify that the combined SDSS and UKIRT spectra are consistent with a typical quasar optical-UV continuum of the form $\text{f}_{\nu} \propto \nu^{-0.5}$ \citep{2001AJ....122..549V}.  By comparing the flux densities at the rest-frame wavelength of 5100 \AA ~to the flux densities at the rest-frame wavelength in the region of 2000 to 3500 \AA , dependent on the redshift, in the SDSS spectrum of each source, we verified that the differences between the two values were within $30\%$, indicating, at most, only modest flux variations.  Such variations, over a temporal baseline of $\sim6$ years in the rest-frame are not atypical for such luminous quasars, even if most of these variations are intrinsic as opposed to measurement errors \citep[see, e.g.,][]{2007ApJ...659..997K}.\\
\subsection{Fitting of the UKIRT Spectra}\label{sec:Fit}
\indent In order to fit the H$\beta$ and H$\alpha$ spectral regions, we used a model consisting of a local, linear continuum, which is a good approximation to a power-law continuum given the relatively narrow spectral band, a broadened \citet{1992ApJS...80..109B} Fe~{\sc ii} emission template, and a multi-Gaussian fit to the emission lines.  The Fe~{\sc ii} template was broadened by a FWHM value that was free to vary between $2000$ and $10000$ km~s$^{-1}$  and, along with the linear continuum, was removed to more accurately fit the H$\beta$ and [O~{\sc iii}] emission lines.  The chosen FWHM to broaden the Fe~{\sc ii} template was determined with a least squares analysis.\\
\indent We fit the H$\beta$ line using two independent Gaussians, constrained by the width and height of the emission line, simultaneously with one Gaussian for each of the [O~{\sc iii}] emission lines.  The Gaussians assigned to the [O~{\sc iii}] emission lines have identical widths and their intensity ratio was fixed to $I($[O~{\sc iii}]~$ \lambda 5007)/I($[O~{\sc iii}]~$\lambda 4959) = 3$.  The wavelengths of the two [O~{\sc iii}] components were fixed to the ratio 5007/4959.  For the available H$\alpha$ features, two Gaussians were fit after a linear continuum was fit and subtracted around the emission line.  We do not detect any [N~{\sc ii}] emission lines while fitting this region, mainly given our low spectral resolution.  The Gaussians were constrained so that the line peak would lie within 1,500 km~s$^{-1}$ from the wavelength that corresponded to the maximum of the emission line region, the widths could range from 0 km~s$^{-1}$ to 15,000 km~s$^{-1}$, and the flux density was restricted to lie within 0 and twice the maximum value of the emission line.\\  
\begin{figure*}
\plotone{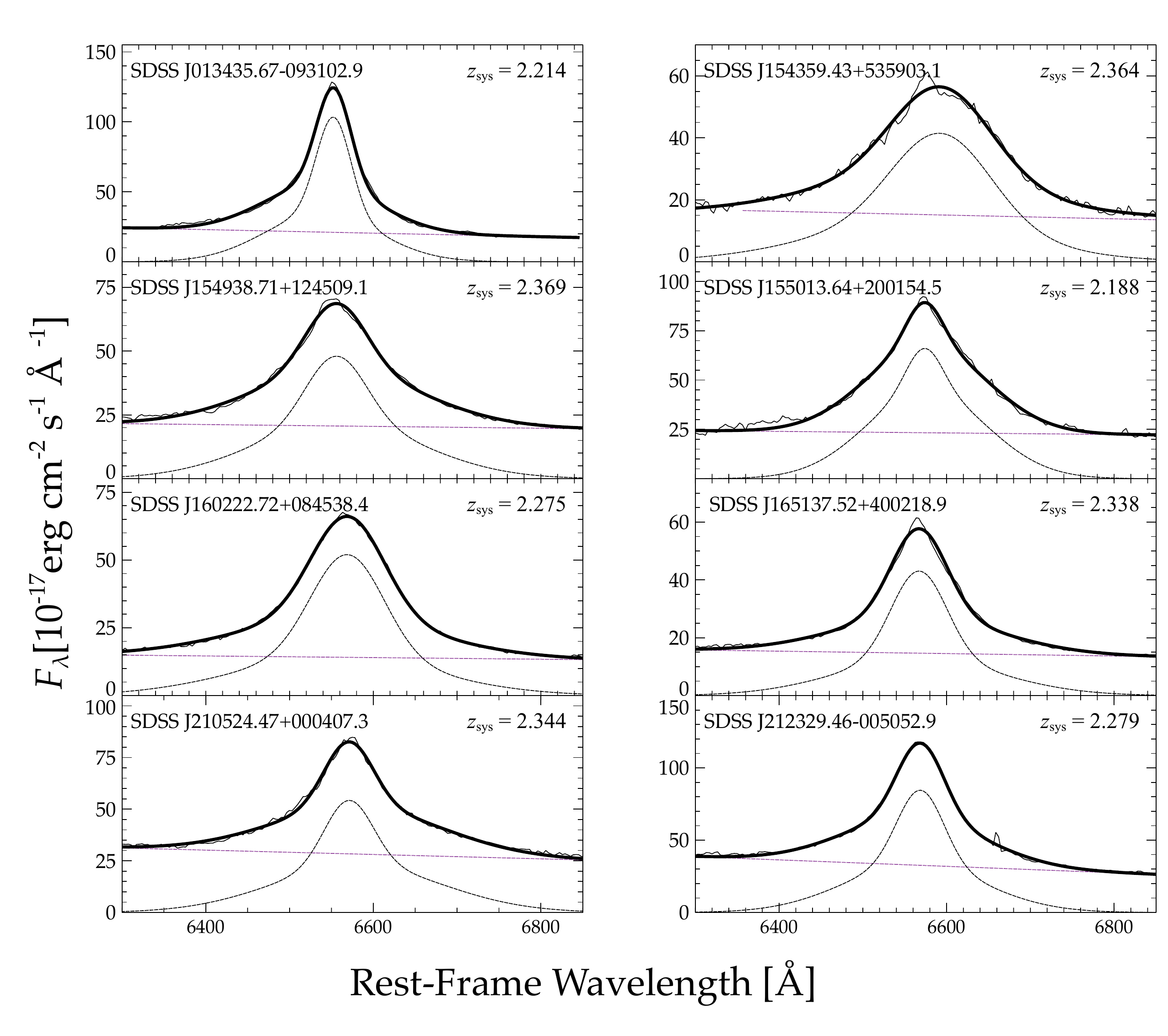}
\caption{NIR spectra of $2.15 < z < 2.65$ quasars.  The spectrum in each panel is given by a thin solid line.  The fit to the H$\alpha$ line and linear continuum are indicated by dashed lines.  The overall fit to each spectrum is given by the bold solid line. \label{fig:ha}}
\end{figure*}
To estimate the uncertainties on the FWHM and rest-frame equivalent width (EW) of the emission lines, we performed the fitting by adjusting the placement of the continuum according to the noise level in the continuum \citep[see, e.g.,][]{2015ApJ...805..124S}.  Namely, by adjusting the local linear continuum between extremes of the noise around each emission line, we were able to derive an estimate for uncertainties on the FWHM and EW values.  For all but two of the sources, the uncertainties on the values of FWHM and EW in the H$\beta$ region are on the order of $\sim5$-$15\%$.  For \hbox{SDSS J014850.64$-$090712.8} and \hbox{SDSS J163300.13$+$362904.8}, these uncertainties are on the order of $\sim40\%$.  Similarly, the uncertainties on the FWHM and EW values for the H$\alpha$ emission line are up to $\sim5\%$.\\
\indent The uncertainties on the wavelengths of the peaks of all the emission lines are up to $\sim300$ km s$^{-1}$.  The majority of this uncertainty arises from the resolution of our spectrograph, however, our choice of a narrow slit was used to combat this.  The uncertainty introduced from the pixel-wavelength calibration is minimal, averaging $\sim5$ km s$^{-1}$.  The narrow [O~{\sc iii}]~$\lambda5007$ emission line provided our most accurate redshift estimates, having uncertainties on wavelength measurements averaging $\sim150$ km s$^{-1}$.  The wavelength uncertainties were determined by evaluating our S/N ratio and repeated measurements of each of the emission lines.\\
\indent Basic spectral properties resulting from those fits are reported in Table~\ref{tab:beta}.  Columns (2), (3), and (4) provide the FWHM, EW, and the observed-frame wavelength of the peak ($\lambda_{\rm peak}$) of the H$\beta$ line, respectively. Columns (5--7) and (8--10) provide similar information for the [O~{\sc iii}]~$\lambda5007$ and H$\alpha$ emission lines, respectively.  The fits for the H$\beta$ and [O~{\sc iii}] emission lines appear in Figure \ref{fig:fit1}, and the fits for the H$\alpha$ emission line appear in Figure \ref{fig:ha}. \\
\subsection{Spectral Fitting of the C~{\sc iv} Emission Lines}\label{sec:c4}
\begin{figure*}
\plotone{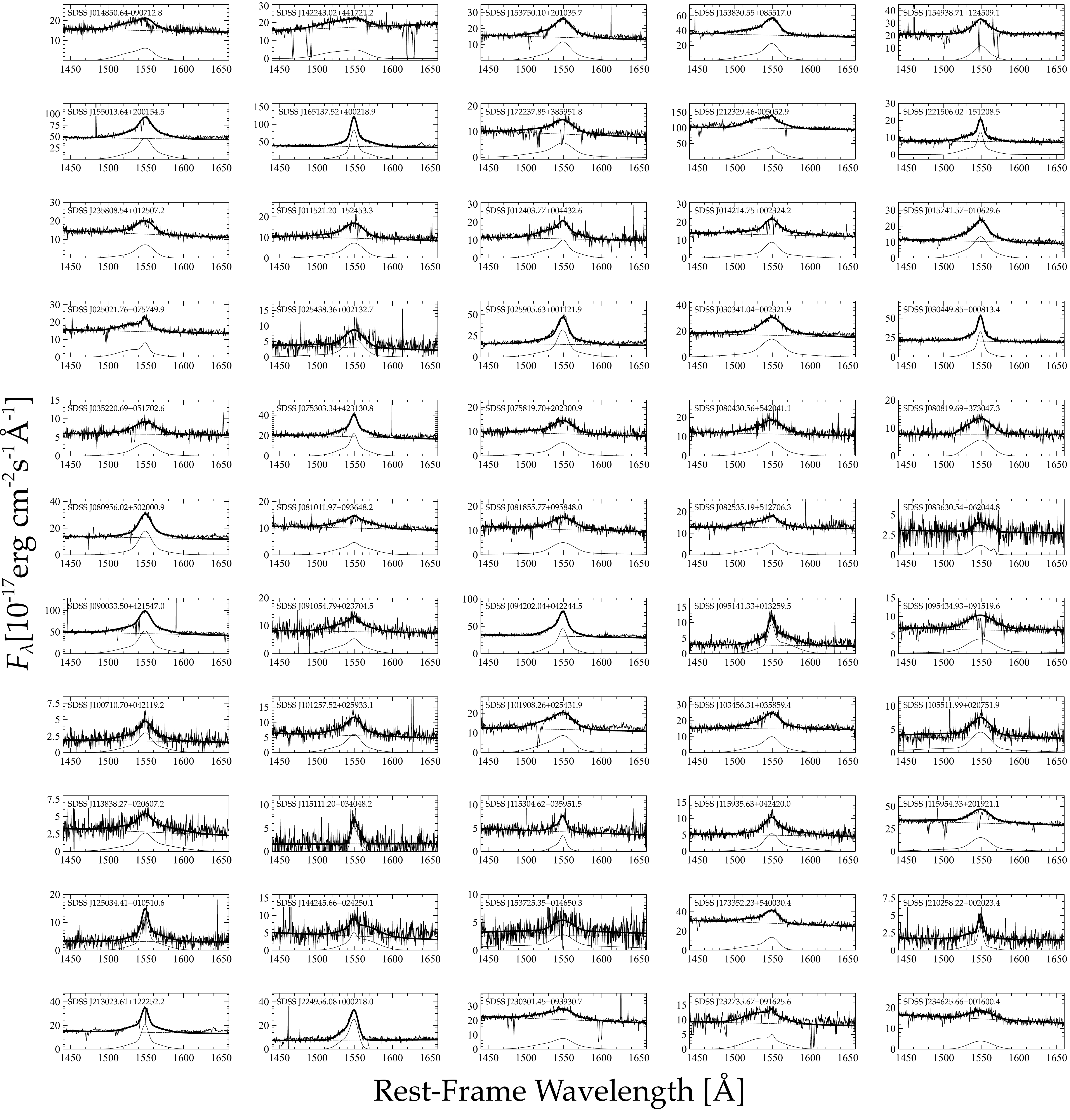}
\caption{C~{\sc iv} fits of all 55 quasars used in the regression analysis.  The spectrum and fit to the C~{\sc iv} emission line in each panel are given by a thin solid line. The linear continuum is indicated by a dashed line. The overall fit to each spectrum is given by the bold solid line.\label{fig:c4}}
\end{figure*}

\indent In order to provide corrections to the UV-based redshifts of our sources, we fit the C~{\sc iv} emission lines present in their SDSS spectra.  These fits appear in Figure \ref{fig:c4}.  As suggested in M17, the parameters needed for the correction of the UV-based redshifts are the FWHM and EW of the C~{\sc iv} line, as well as the monochromatic luminosity of the continuum at a rest-frame wavelength of $1350$ \AA.\\
\indent The C~{\sc iv} emission line was fit with a local, linear continuum and two independent Gaussians under the same constraints as we report for the H$\beta$ and H$\alpha$ emission lines.  The spectral properties resulting from this fitting procedure are reported in Table \ref{tab:c4}.  The uncertainties in each of these measurements were determined by the same method used when evaluating the rest-frame optical emission line uncertainties.  Along with this fit, the continuum luminosity, $L_{1350}$, has also been derived by measuring the continuum flux density at rest-frame $\lambda 1350$ \AA\ and employing our chosen cosmology.  These values also appear in Table \ref{tab:c4}.  

\begin{figure}
\plotone{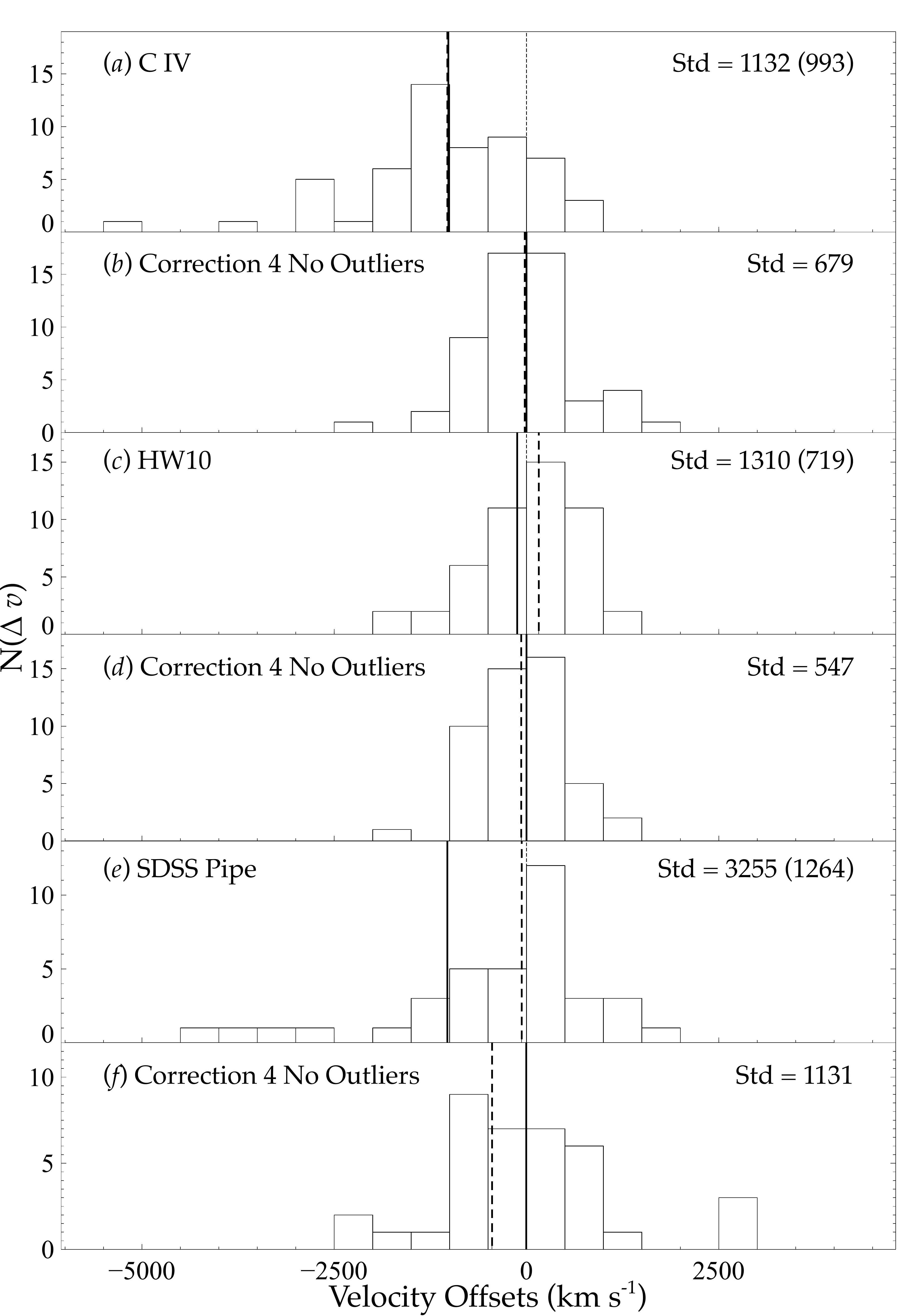}
\caption{Velocity offsets relative to $z_{\rm sys}$ before (panels \textit{a}, \textit{c}, and \textit{e}) and after (panels \textit{b}, \textit{d}, and \textit{f}) the correction provided in bold face in Table \ref{tab:stats}.  The numbers reported in parentheses are the standard deviations of the original distributions without the outliers.  The mean (solid line) and median (dashed line) are marked in each panel.  SDSS J142243.02$+$441721.2 does not appear on the SDSS Pipe panel, for clarity, because of its abnormally large velocity offset. The outliers that were removed are discussed in Section \ref{sec:4}. \label{fig:voff}}
\end{figure}

\begin{figure*}
\plotone{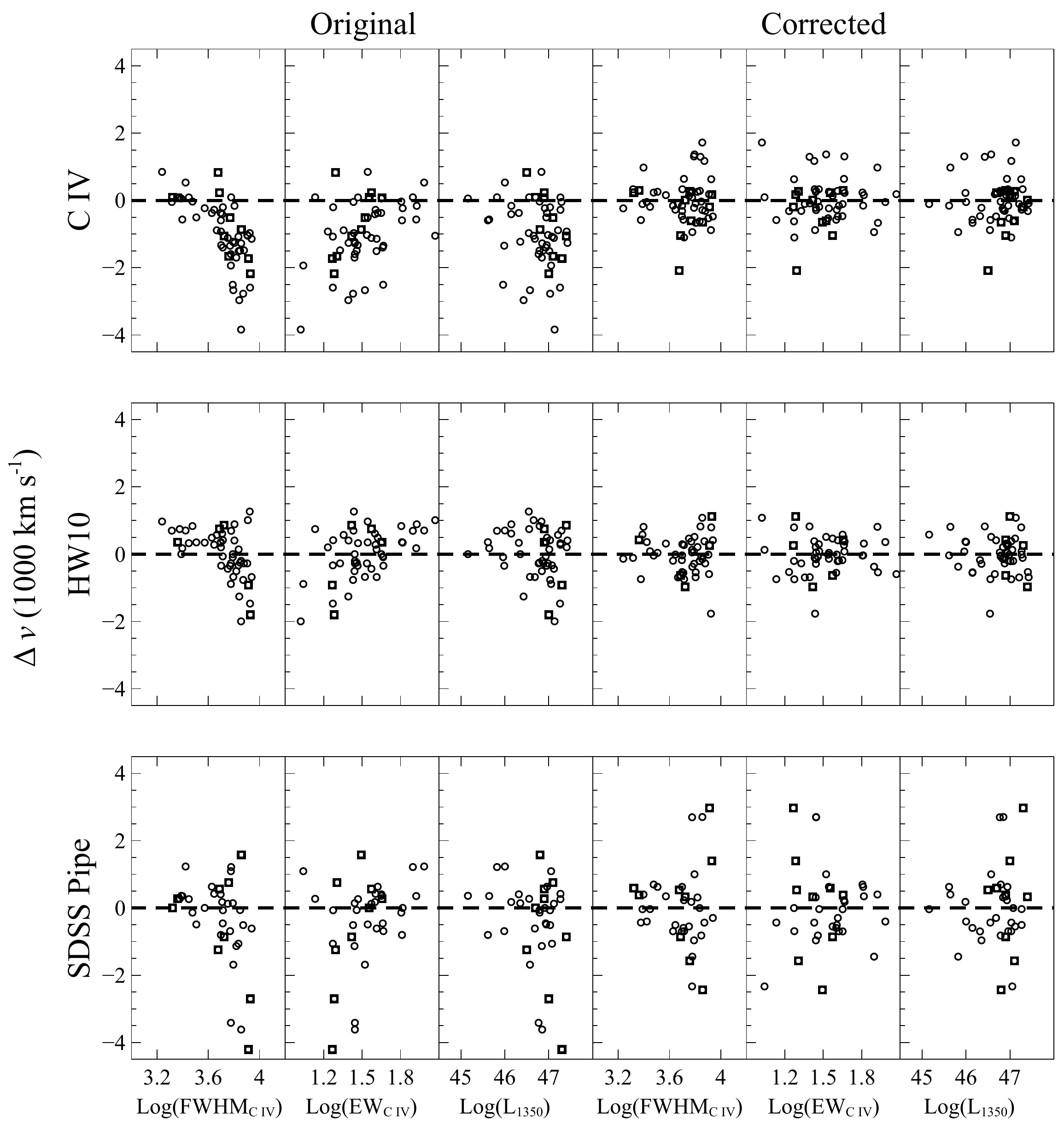}
\caption{The residual velocity offsets with respect to $z_{\rm sys}$ before, three leftmost panels, and after, three rightmost panels, correction are presented against our regression parameters.  The corrected method displayed refers to Correction 4 without outliers reported in Table \ref{tab:stats}.  Squares (circles) represent data from M17 (UKIRT; this work).  The outliers discussed in Section \ref{sec:out} do not appear in this plot given their abnormally large velocity offsets. \label{fig:res}}
\end{figure*}

\begin{deluxetable*}{lccccccc}
\tablecaption{Spectral Measurements of C~{\sc iv} \label{tab:c4}}
\tablewidth{0pt}
\tablehead{
\colhead{} & \colhead{FWHM\tablenotemark{a}} &
\colhead{EW\tablenotemark{a}} & \colhead{log $L_{1350}$\tablenotemark{a}} & \colhead{FWHM\tablenotemark{b}} &
\colhead{EW\tablenotemark{b}} & \colhead{log $L_{1350}$\tablenotemark{b}} & \colhead{$\lambda_{\rm peak}$\tablenotemark{b}}\\
\colhead{Quasar Name} & \colhead{(km s$^{-1}$)}  & \colhead{(\AA)} & \colhead{(erg s$^{-1}$)} & \colhead{(km s$^{-1}$)} & \colhead{(\AA)} & \colhead{(erg s$^{-1}$)} & \colhead{{(\AA)}} 
}
\startdata
 SDSS J013435.67$-$093102.9 & 1045 & \nodata & \nodata & 
 \nodata & \nodata & \nodata & \nodata \\
 SDSS J014850.64$-$090712.8 & 9545 & 16.3  & 47.0 & 
 8490 & 19.2 & 47.0 & 6657\\
 SDSS J073607.63$+$220758.9\tablenotemark{c} & \nodata & \nodata & \nodata & 
 2496 & 10.0 & 46.8 & 6872 \\
 SDSS J142243.02$+$441721.2 & 12475 & 20.8 & 47.0 & 
 12326 & 17.9 & 47.0 & 7082\\
 SDSS J153750.10$+$201035.7 & 6080 & 37.9 & 47.1 &
 5886 & 33.3 & 47.1 & 6824\\
 SDSS J153830.55$+$085517.0 & 5754 & 27.1 & 47.5 &
 5279 & 26.2 & 47.4 & 7023 \\
 SDSS J154359.43$+$535903.1\tablenotemark{c} & 4713 & 42.6 & 47.0 &
 4553 & 36.9 & 46.9 & 5211\\
 SDSS J154446.33$+$412035.7\tablenotemark{c} & 15266 & 192.3 & 46.3 &
 7350 & 34.4 & 46.6 & 7001\\
 SDSS J154938.71$+$124509.1 & 4207 & 24.2 & 46.6 &
 4740 & 19.6 & 46.5 & 5233\\
 SDSS J155013.64$+$200154.5 & 4273 & 42.6 & 47.0 &
 4858 & 37.4 & 46.9 & 4942\\
 SDSS J160222.72$+$084538.4\tablenotemark{c} & 4150 & 27.8 & 47.0 &
 5615 & 30.7 & 47.0 & 5065\\
 SDSS J163300.13$+$362904.8\tablenotemark{c} & 6963 & 34.9 & 46.9 &
 6614 & 42.0 & 46.8 & 7067\\
 SDSS J165137.52$+$400218.9 & 2818 & 49.9 & 46.9 &
 2297 & 45.2 & 46.9 & 5172\\
 SDSS J172237.85$+$385951.8 & \nodata & \nodata & \nodata &
 7208 & 31.1 & 46.8 & 6745 \\
 SDSS J210524.47$+$000407.3\tablenotemark{c} & 12603 & 36.4 & 47.1 &
 7990 & 11.9 & 46.8 & 5098\\
 SDSS J212329.46$-$005052.9 & 8549 & 16.2 & 47.4 &
 8168 & 18.5 & 47.3 & 5050\\
 SDSS J221506.02$+$151208.5 & \nodata & \nodata & \nodata & 
 2094 & 35.8 & 46.7 & 6638 \\
 SDSS J235808.54$+$012507.2 & \nodata & \nodata & \nodata &
 5728 & 20.2 & 47.1 & 6761 
\enddata
\tablenotetext{a}{Columns (2), (3), and (4) were reported in \citet{2011ApJS..194...45S}.}
\tablenotetext{b}{Columns (5), (6), (7), and (8) were measured from SDSS spectra, as described in the text.}
\tablenotetext{c}{Indicates broad absorption around the C~{\sc iv} line.}
\end{deluxetable*}

\section{Results} \label{sec:3}
Combined with the sources in M17, we have a total of 63 objects in our sample, of which, six of our UKIRT objects were excluded from further analysis due to broad absorption line (BAL)\footnote{Five of these sources are based on BAL quasar identification from \citet{2011ApJS..194...45S}; SDSS J073607.63+220758.9 was identified as a BAL quasar following our visual inspection of its SDSS spectrum.} identification: these are noted in Table \ref{tab:log}.  We then remove an additional BAL quasar, \hbox{SDSS J014049.18$-$083942.5}, from the sample in M17.  Furthermore, we have excluded SDSS J013435.67$-$093102.9 from our sample given that it is a lensed quasar and its rest-frame UV spectrum is severely attenuated by the foreground lensing galaxy \citep[see e.g.,][]{2006ApJ...641...70O}.  Measurements of the C~{\sc iv} emission line for 52 out of the 55 sources in our combined sample are available in \citet{2011ApJS..194...45S}.  The C~{\sc iv} FWHM and EW measurements we obtained for 40 of these sources agree to within $\sim 20\%$ with those from \citet{2011ApJS..194...45S}; similarly, measurements of 10 of these sources agree to within $\sim 65\%$.  Generally, these discrepancies are inversely proportional to the signal-to-noise ratios of the SDSS spectra and are larger in the presence of narrow absorption lines.  The spectra for \hbox{SDSS J025438.36$+$002132.7}  and \hbox{SDSS J153725.35$-$014650.3} had extremely poor signal-to-noise ratios, resulting in discrepancies of $108\%$ and $53\%$ for FWHM, and $57\%$ and $210\%$ for EW, respectively, between our measured values and the ones reported in \citet{2011ApJS..194...45S}.  Substituting our values with the ones reported in \citet{2011ApJS..194...45S} for these objects did not have a significant impact on further analysis.\\
\indent The observed-frame wavelength of the peak of the C~{\sc iv} emission line was compared to the value predicted by the systemic redshift ($z_{\rm sys}$) to determine the velocity offset of this line.  We determine $z_{\rm sys}$ from the line peak of the emission line with the smallest measurement uncertainty.  In order, we take our systemic redshift from [O~{\sc iii}] ($\sim50$ km s$^{-1}$), Mg~{\sc ii} ($\sim200$ km s$^{-1}$) and H$\beta$ ($\sim400$ km s$^{-1}$) \citep{2016ApJ...831....7S}.  The C~{\sc iv} velocity offsets are shown and reported in Figure \ref{fig:voff} and Table \ref{tab:red}, respectively.  In Table \ref{tab:red}, we also report the redshift measurements provided for these sources in HW10 and \citet[][hereafter P18]{2018A&A...613A..51P}, where applicable.  The velocity offsets introduced from these redshifts with respect to $z_{\rm sys}$ are presented in Figure \ref{fig:voff} and Table \ref{tab:red}.  In addition to the velocity offsets for the sources in our UKIRT sample, the velocity offsets from Table 1 of M17 have been included in the following regression analysis.  The C~{\sc iv} emission line properties for the M17 sample are reported in Table \ref{tab:Mason}.\\
\indent We note that the $\Delta v_{\text{C~{\sc iv}}}$ values used in M17 differ from the $\Delta v_{\text{C~{\sc iv}}}$ values we compute for the M17 sample since M17 used the $\Delta v_{\text{C~{\sc iv}}}$ values from \citet{2011ApJS..194...45S}, combined with the redshift determined from the SDSS Pipeline, in order to find $z_{\text{C~{\sc iv}}}$. Our $\Delta v_{\text{C~{\sc iv}}}$ values follow directly from the measurement of $\lambda_{\rm peak}$ (C~{\sc iv}) and our derived $z_{\rm sys}$.  The origin of the discrepancies between the two velocity offsets used stems from the uncertainty in the $\Delta v_{\text{C~{\sc iv}}}$ values discussed in \citet{2011ApJS..194...45S}.  The differences between the $\Delta v_{\text{C~{\sc iv}}}$ values we use and those used by M17 are rather small, and using the latter values do not change our results significantly.\\
\indent  A multiple regression analysis has been performed on the velocity offsets and the C~{\sc iv} emission line properties such that:
\begin{equation}\label{eq:coeff}
\begin{split}
\Delta v~(\text{km}~\text{s}^{-1})= \alpha \log_{10}(\text{FWHM}_{\text{C~{\sc iv}}}) \\ +  \beta \log_{10}(\text{EW}_{\text{C~{\sc iv}}}) +  \gamma \log_{10}(L_{1350})
\end{split}
\end{equation}
where $\Delta v$ is the velocity offset and $\alpha$, $\beta$, and $\gamma$ are the coefficients associated with our regression analysis. The velocity offset created by each redshift derivation method was determined by the following equation
\begin{equation}
\begin{split}
\Delta v = c \bigg( \frac{z_{\rm meas} - z_{\rm sys}}{1 + z_{\rm sys}} \bigg).
\end{split}
\end{equation}
Where $z_{\rm meas}$ is the redshift derived using various methods and reported in the studies indicated below. In order to derive the most reliable redshift correction, four regressions were performed using the following parameters from Equation \ref{eq:coeff}:
\begin{enumerate}
\item $\log_{10}(\text{FWHM}_{\text{C~{\sc iv}}})$, $\log_{10}(\text{EW}_{\text{C~{\sc iv}}})$
\item $\log_{10}(\text{FWHM}_{\text{C~{\sc iv}}})$, $\log_{10}(L_{1350})$
\item $\log_{10}(\text{EW}_{\text{C~{\sc iv}}})$, $\log_{10}(L_{1350})$
\item All three parameters
\end{enumerate}
In total, this regression analysis was performed on redshifts determined from: 1) the measured line peak of the C~{\sc iv} emission line, 2) HW10, and 3) the SDSS Pipeline.  The coefficients, errors, and confidence statistics from Equation \ref{eq:coeff}, determined in each of these cases, are reported in Table \ref{tab:coeff1}.  For the confidence statistics, we report the $t$-Value \citep[e.g.,][]{sheskin07} to determine the importance of each individual parameter.\\
\indent The residuals of the velocity offsets after each correction has been determined have been analyzed, and basic statistics resulting from these residuals are listed in Table \ref{tab:stats}.  The residuals before and after correction are presented in Figure \ref{fig:res}.  The residual distributions show the significant reduction in the velocity offsets before and after each correction.  The corrected velocity offsets for C~{\sc iv}- and HW10-based redshifts are closer to zero than the corrected velocity offsets for the SDSS Pipeline-based redshift, representative of the larger $\sigma$ value associated with SDSS Pipeline redshift estimates.  From evaluating the best fitting coefficients and statistics reported for each correction, we determined the correction that we consider to provide the most reliable results.  This correction has been emphasized in bold face in the text.\\  
\subsection{SDSS J142243.02$+$441721.2 and \hbox{SDSS J115954.33$+$201921.1}}\label{sec:out}
\indent SDSS J142243.02$+$441721.2 from our UKIRT sample has significantly larger velocity offsets compared to the rest of the combined sample.  The velocity offsets determined from C~{\sc iv}, HW10, and SDSS Pipeline are \hbox{$-5097$} km s$^{-1}$, \hbox{$-7740$} km s$^{-1}$, and \hbox{$-16384$} km s$^{-1}$, respectively.  The latter velocity offset stems from a misidentification of spectral features in the SDSS spectrum of the source as manifested by the SDSS Pipeline products.  The SDSS Pipeline redshift for this source is $z = 3.396$ while the SDSS Visual Inspection value is $z = 3.615$.  The disparity between these estimates confirms the misidentification of the emission lines by the SDSS Pipeline.  Because the velocity offsets for this source had a significant impact on the regression analysis and may be misleading, we have provided the results of the regression analysis with and without this object in Table \ref{tab:stats}.\\
\indent The velocity offset of SDSS J115954.33$+$201921.1 from the M17 sample, with respect to the redshift determined by the SDSS Pipeline is $-10642$ km s$^{-1}$, which is significantly larger than the respective values of the combined sample, excluding SDSS J142243.02$+$441721.2.  \hbox{SDSS J115954.33$+$201921.1} was also removed from the SDSS Pipeline regression as discussed further in Section \ref{sec:4}.  Here too, the disparity between the SDSS Pipeline redshift value ($z = 3.330$) and the respective Visual Inspection value ($z = 3.425$) indicates a misidentification of spectral features by the SDSS Pipeline. 

\begin{deluxetable*}{lcccccc}
\tablecaption{Redshift Comparison \label{tab:red}}
\tablewidth{0pt}
\tablehead{
\colhead{} & 
 \colhead{} & \colhead{$\Delta v$} & 
 \colhead{} &  \colhead{$\Delta v$} &
 \colhead{} &  \colhead{$\Delta v$} \\
 \colhead{Quasar} & 
 \colhead{$z_{\rm \text{C~{\sc iv}}}\tablenotemark{a}$} & \colhead{(km s$^{-1}$)} & 
\colhead{$z_{\rm pipe}\tablenotemark{b}$} & \colhead{(km s$^{-1}$)} &
 \colhead{$z_{\rm HW10}\tablenotemark{c}$} &  \colhead{(km s$^{-1}$)}
} 
\startdata
 SDSS J013435.67$-$093102.9 & 2.214 & \nodata &
  \nodata & \nodata &
   2.225 & 1029\\
 SDSS J014850.64$-$090712.8 & 3.274 & -3786 &
   3.290 & -2691 & 
  3.303 & -1796\\
 SDSS J073607.63$+$220758.9 & 3.436 & -607 & 
 3.464 & 1285 & 
  \nodata  & \nodata \\
 SDSS J142243.02$+$441721.2 & 3.572 & -5097 & 
  3.397 & -16384 &
  3.531 & -7740\\
 SDSS J153750.10$+$201035.7 & 3.405 & -544 & 
  \nodata  & \nodata &
  \nodata & \nodata\\
 SDSS J153830.55$+$085517.0 & 3.535 & -989 & 
  3.537 & -856 & 
  3.563 & 858\\
 SDSS J154359.43$+$535903.1 & 2.365 & 89 & 
  2.370 & 536 &
  2.379 & 1341\\
 SDSS J154446.33$+$412035.7 & 3.520 & -3087 & 
  3.569 & 131  &
  3.551 & -1049\\
 SDSS J154938.71$+$124509.1 & 2.378 & 801 & 
 2.355 & -1244 & 
  \nodata & \nodata \\
 SDSS J155013.64$+$200154.5 & 2.190 & 188 & 
  2.194 & 565 &
  2.196 & 754\\
 SDSS J160222.72$+$084538.4 & 2.270 & -458 & 
 \nodata  & \nodata &
  2.276 & 92\\
 SDSS J163300.13$+$362904.8 & 3.562 & -525 & 
 3.538 & -2093 &
  3.575 & 328\\
 SDSS J165137.52$+$400218.9 & 2.339 & 90 & 
  2.341 & 270 & 
  2.342 & 360\\
 SDSS J172237.85$+$385951.8 & 3.350 & -1168 & 
3.390 & 1584 & 
  \nodata & \nodata \\
 SDSS J210524.47$+$000407.3 & 2.293 & -4575 & 
 \nodata  & \nodata & 
 2.307 & -3301\\
 SDSS J212329.46$-$005052.9 & 2.255 & -1376 & 
  2.233 & -3395 &
  2.269 & -92\\
 SDSS J221506.02$+$151208.5 & 3.285 & 70 & 
  3.284 & 0 & 
 \nodata & \nodata \\
 SDSS J235808.54$+$012507.2 & 3.366 & -1572 & 
  3.400 & 753 &
  \nodata & \nodata 
\enddata
\tablenotetext{a}{Redshifts determined from the $\lambda_{\rm peak}$ reported in Column (8) of Table \ref{tab:c4}.}
\tablenotetext{b}{Acquired from P18.}
\tablenotetext{c}{Acquired from HW10}
\end{deluxetable*}

\begin{deluxetable*}{lcccccc}
\tablecaption{C~{\sc iv} Spectral Properties of the M17 Sample\label{tab:Mason}}
\tablewidth{0pt}
\tablehead{ \colhead{}  & \colhead{} & \colhead{} & \colhead{$\text{FWHM}$} & \colhead{$\text{EW}$} & \colhead{log $L_{1350}$} & \colhead{$\lambda_{\rm peak}$}\\
\colhead{Quasar}  & \colhead{$z_{\rm HW10}$} & \colhead{$z_{\rm pipe}$} & \colhead{(km s$^{-1}$)} & \colhead{(\AA)} & \colhead{(erg s$^{-1}$)} & \colhead{(\AA)}
}
\startdata
 SDSS J011521.20$+$152453.3 & 3.433 & 3.418 & 
 6236 & 33.3 & 46.6 & 6821 \\
 SDSS J012403.77$+$004432.6 & 3.827 & 3.836 &
 5646 & 37.4 & 47.1 & 7460\\
 SDSS J014049.18$-$083942.5\tablenotemark{a} & 3.726 & \nodata &
 4635 & 22.7 & 47.2 & 7285\\
 SDSS J014214.75$+$002324.2 & 3.374 & \nodata & 
 5013 & 29.2 & 47.0 & 6753\\
 SDSS J015741.57$-$010629.6 & 3.571 & 3.565 &
 5158 & 45.9 & 46.9 & 7049 \\
 SDSS J025021.76$-$075749.9  & 3.344 & 3.337 & 
 5173 & 18.8 & 47.0 & 6715\\
 SDSS J025438.36$+$002132.7  & 2.464 & 2.470 & 
 5998 & 78.8 & 45.8 & 5355\\
 SDSS J025905.63$+$001121.9  & 3.377 & 3.372 & 
 3728 & 65.6 & 46.9 & 6767\\
 SDSS J030341.04$-$002321.9 & 3.235 & \nodata &
 6865 & 41.0 & 47.0 & 6524\\
 SDSS J030449.85$-$000813.4  & 3.296 & \nodata &
 2066 & 27.1 & 47.3 & 6638\\
 SDSS J035220.69$-$051702.6  & 3.271 & \nodata &
 6939 & 24.7 & 46.4 & 6578\\
 SDSS J075303.34$+$423130.8  & 3.595 & 3.594 &
 2804 & 29.4 & 47.3 & 7112\\
 SDSS J075819.70$+$202300.9  & 3.753 & 3.743 &
 6583 & 27.6 & 46.8 & 7333\\
 SDSS J080430.56$+$542041.1  & 3.755 & 3.758 &
 7047 & 28.7 & 46.8 & 7335\\
 SDSS J080819.69$+$373047.3 & 3.477 & 3.426 &
 7183 & 27.8 & 46.9 & 6910\\
 SDSS J080956.02$+$502000.9 & 3.288 & 3.290 &
 4240 & 41.9 & 47.0 & 6623\\
 SDSS J081011.97$+$093648.2 & 3.387 & \nodata &
7558 & 21.3 & 46.9 & 6768\\
 SDSS J081855.77$+$095848.0 & 3.688 & 3.692 &
 7446 & 26.9 & 47.0 & 7213\\
 SDSS J082535.19$+$512706.3 & 3.507 & 3.496 &
 6839 & 18.7 & 47.1 & 6964 \\
 SDSS J083630.54$+$062044.8 & 3.387 & 3.413 &
 5971 & 11.0 & 47.1 & 6767\\
 SDSS J090033.50$+$421547.0  & 3.294 & 3.296 &
 4421 & 40.3 & 47.3 & 6639\\
 SDSS J091054.79$+$023704.5 & 3.290 & 3.292 &
 6184 & 27.7 & 46.4 & 6618\\
 SDSS J094202.04$+$042244.5 & 3.284 & 3.272 &
 3208 & 35.0 & 46.9 & 6617\\
 SDSS J095141.33$+$013259.5  & 2.419 & 2.425 &
 2645 & 96.5 & 46.0 & 5293\\
 SDSS J095434.93$+$091519.6 & 3.398 & 3.399 &
 8671 & 41.1 & 46.7 & 6802\\
 SDSS J100710.70$+$042119.2 & 2.367 & 2.354 &
 4988 & 64.8 & 45.6  & 5199\\
 SDSS J101257.52$+$025933.1 & 2.441 & 2.436 &
 5106 & 39.9 & 46.1 & 5312\\
 SDSS J101908.26$+$025431.9  & 3.379 & \nodata &
 8012 & 34.5 & 47.0 & 6766\\
 SDSS J103456.31$+$035859.4 & 3.388 & 3.342 &
  5972 & 27.8  & 46.8 & 6767\\
 SDSS J105511.99$+$020751.9 & 3.404 & \nodata &
 6372 & 84.5 & 46.1 & 6798\\
 SDSS J113838.27$-$020607.2 & 3.347 & 3.342 &
 5888 & 46.4 & 46.0& 6711\\
 SDSS J115111.20$+$034048.2 & 2.337 & 2.341 &
 2448 & 44.8 & 45.2 & 5170\\
 SDSS J115304.62$+$035951.5  & 3.437 & 3.430 &
2379 & 13.6 & 46.6 & 6858\\
 SDSS J115935.63$+$042420.0 & 3.456 & 3.457 &
 4969 & 44.8 & 46.3 & 6886\\
 SDSS J115954.33$+$201921.1 & 3.432 & 3.269 &
 6360 & 24.8 & 47.4 & 6827\\
 SDSS J125034.41$-$010510.6 & 2.399 & 2.401 &
 2494 & 83.7 & 45.6 & 5252\\
 SDSS J144245.66$-$024250.1 & 2.355 & \nodata &
 6176 & 46.2 & 46.0 & 5155\\
 SDSS J153725.35$-$014650.3 & 3.467 & \nodata &
 8098 & 117.7 & 46.7 & 6872\\
 SDSS J173352.23$+$540030.4 & 3.435 & \nodata &
 4994 & 17.1 & 47.4 & 6844\\
 SDSS J210258.22$+$002023.4 & 3.342 & \nodata &
 1733 & 35.0 & 46.8 & 6723\\
 SDSS J213023.61$+$122252.2 & 3.279 & \nodata &
 2596 & 33.6 & 47.0 & 6615\\
 SDSS J224956.08$+$000218.0 & 3.323 & 3.309 &
 2994 & 64.0  & 46.8 & 6677\\
 SDSS J230301.45$-$093930.7 & 3.470 & \nodata &
 8425 & 18.7 &  47.3 & 6898\\
 SDSS J232735.67$-$091625.6 & 3.470 & \nodata &
 8378 & 27.3 & 46.5 & 6582\\
 SDSS J234625.66$-$001600.4 & 3.281 & \nodata & 
 7172 & 10.5 & 47.1 & 6892
\enddata
\tablenotetext{a}{This object was excluded from the regression analysis after visually inspecting its SDSS spectrum and determining it was a BAL quasar.}
\tablecomments{The $z_{\rm sys}$ values used in determining the velocity offsets are reported in Column 3 of Table 1 in M17.}
\end{deluxetable*}

\begin{deluxetable*}{lccccc}
\tablecaption{Correction Coefficients \label{tab:coeff1}}
\tablewidth{0pt}
\tablehead{
\colhead{Correction} & \colhead{Equation} & \colhead{Coefficients} & \colhead{Value} & \colhead{Error} & \colhead{$t$-Value}
}
\startdata
C~{\sc iv} & $\alpha \log_{10}(\text{FWHM}_{\text{C~{\sc iv}}}) +  \beta \log_{10}(\text{EW}_{\text{C~{\sc iv}}})$ & $\alpha$ & -1301 & 195 & -6.68 \\
  & & $\beta$ & 2501 & 472 & 5.29\\
 \hline
 & $\alpha \log_{10}(\text{FWHM}_{\text{C~{\sc iv}}}) +  \gamma \log_{10}(L_{1350})$ & $\alpha$ & -3966 & 600 & -6.61\\
 & & $\gamma$ & 293 & 48 & 6.14 \\
 \hline
 & $\beta \log_{10}(\text{EW}_{\text{C~{\sc iv}}}) + \gamma \log_{10}(L_{1350})$ & $\beta$ & 2058 & 601 & 3.43 \\
 & & $\gamma$ & -88 & 20 & -4.50\\
 \hline
 & $\alpha \log_{10}(\text{FWHM}_{\text{C~{\sc iv}}}) +  \beta \log_{10}(\text{EW}_{\text{C~{\sc iv}}}) +  \gamma \log_{10}(L_{1350})$ & $\alpha$ & -3670 & 549 & -6.68 \\
 & & $\beta$ & 1604 & 450 & 3.57 \\
 & & $\gamma$ & 217 & 48 & 4.53 \\
\hline\hline
HW10 & $\alpha \log_{10}(\text{FWHM}_{\text{C~{\sc iv}}}) +  \beta \log_{10}(\text{EW}_{\text{C~{\sc iv}}})$ & $\alpha$ & -1069 & 254 & -4.22 \\
 & & $\beta$ & 2517 & 612 & 4.11 \\
 \hline
  & $\alpha \log_{10}(\text{FWHM}_{\text{C~{\sc iv}}}) +  \gamma \log_{10}(L_{1350})$ & $\alpha$ & -3191 & 869 & -3.67 \\
 & & $\gamma$ & 251 & 69 & 3.63\\
   \hline
 & $\beta \log_{10}(\text{EW}_{\text{C~{\sc iv}}}) +  \gamma \log_{10}(L_{1350})$ & $\beta$ & 2219 & 715 & 3.10 \\
 & & $\gamma$ & -75 & 24 & -3.18\\
   \hline
 & $\alpha \log_{10}(\text{FWHM}_{\text{C~{\sc iv}}}) +  \beta \log_{10}(\text{EW}_{\text{C~{\sc iv}}}) +  \gamma \log_{10}(L_{1350})$ & $\alpha$ & -2834 & 819 & -3.46 \\
 & & $\beta$ & 1877 & 652 & 2.88 \\
 & & $\gamma$ & 161 & 71 & 2.26 \\
\hline\hline
SDSS Pipe & $\alpha \log_{10}(\text{FWHM}_{\text{C~{\sc iv}}}) +  \beta \log_{10}(\text{EW}_{\text{C~{\sc iv}}})$ & $\alpha$ & -2380 & 785 & -3.03 \\
 & & $\beta$ & 5087 & 1891 & 2.69\\
 \hline
  & $\alpha \log_{10}(\text{FWHM}_{\text{C~{\sc iv}}}) +  \gamma \log_{10}(L_{1350})$ & $\alpha$ & -8024 & 2732 & -2.94 \\
 & & $\gamma$ & 613 & 216 & 2.83\\
   \hline
 & $\beta \log_{10}(\text{EW}_{\text{C~{\sc iv}}}) +  \gamma \log_{10}(L_{1350})$ & $\beta$ & 4732 & 2240 & 2.11 \\
 & & $\gamma$ & -176 & 74 & -2.39\\
   \hline
 & $\alpha \log_{10}(\text{FWHM}_{\text{C~{\sc iv}}}) +  \beta \log_{10}(\text{EW}_{\text{C~{\sc iv}}}) +  \gamma \log_{10}(L_{1350})$ & $\alpha$ & -6814 & 2830 & -2.41\\
 & & $\beta$ & 3114 & 2212 & 1.41 \\
 & & $\gamma$ & 416 & 255 & 1.63
\enddata
\end{deluxetable*}

\section{Discussion} \label{sec:4}
The results of our multiple regression analysis indicate that the most reliable redshift is obtained by  correcting the HW10-based redshift employing the FWHM and EW of the C IV line, the monochromatic luminosity at rest-frame 1350 \AA, and the respective coefficients listed under the fourth correction to the HW10 method from Table \ref{tab:coeff1}.  Using this correction, and removing SDSS J142243.02$+$441721.2 from the analysis (see Sec. \ref{sec:out}), we were able to reduce the uncertainty on the redshift determination from 731 km s$^{-1}$ to 543 km s$^{-1}$, yielding an improvement of $\sim25\%$ with respect to the HW10-based redshifts; similarly, the mean systematic offset of the redshift determination is reduced from $-137$ km s$^{-1}$ to $+1$ km s$^{-1}$ (see Table \ref{tab:stats}).  For comparison, utilizing only the M17 sample of 44 sources, the uncertainty on the HW10-based redshifts is reduced by $\sim20\%$. The addition of the five sources from our UKIRT sample that have HW10-based redshifts,  comprising a $\sim10\%$ increase in the number of sources with respect to the M17 sample, therefore helped to further reduce the uncertainty on the HW10-based redshifts from $\sim20\%$ to $\sim25\%$.  We anticipate that by utilizing a more representative of several hundred high-redshift quasars, we will be able to further improve these uncertainties significantly and the results will become increasingly less biased to small number statistics (e.g., Matthews et al., in prep.).\\
\indent  We note that, when we include the source with the highly discrepant $\Delta v_{\text{C~{\sc iv}}}$ value, \hbox{SDSS J142243.02$+$441721.2}, in the regression analysis, the best redshift estimates are obtained from the corrected C~{\sc iv}-based redshifts (see Table \ref{tab:stats}). In this case, the mean systematic redshift offsets reduces from $-1023$ km s$^{-1}$ to $-8$ km s$^{-1}$ and the uncertainty on the redshifts determination decreases from 1135 km s$^{-1}$ to 746 km s$^{-1}$ (a $\sim34\%$ improvement).\\
\indent As it is apparent, even with this sample of 55 quasars, the methods to determine redshift using rest-frame UV features provide uncertainties as large as $\approx 500 - 700$ km s$^{-1}$.  As reported in the first row of each section of Table \ref{tab:stats}, the uncorrected redshift determinations are significantly inaccurate and imprecise.  C~{\sc iv}-based redshifts have a mean systematic offset of $\sim1000$ km/s (a blueshift) and a similar value for $\sigma$ (the standard deviation). The HW10 method further improves these C~{\sc iv}-based redshifts by reducing the systematic offsets by $\sim900$ km s$^{-1}$ and $\sigma$ by $\sim300$ km s$^{-1}$. Our prescription further reduces the systematic offset by an additional $\sim100$ km s$^{-1}$ and reduces $\sigma$ by an additional $\sim200-300$ km s$^{-1}$.  Using the SDSS Pipeline redshift estimate, determined from a principal component analysis on multiple features of a spectrum simultaneously \citep[e.g.,][]{2012AJ....144..144B}, the mean systematic velocity offset for our combined sample is the largest and extends beyond $1000$ km s$^{-1}$ with a standard deviation of $1324$ km s$^{-1}$.  Overall, albeit utilizing a smaller combined sample with respect to the samples we use for C~{\sc iv}- and HW10-based redshifts, the redshifts determined from the SDSS Pipeline provide the least reliable results (see Table \ref{tab:stats}). Our best correction applied to these redshifts improves the mean systematic velocity offset by $\sim1000$ km s$^{-1}$, similar to the improvement achieved for C~{\sc iv}-based redshifts, but yields only a modest improvement in $\sigma$ which remains large.\\
\indent In order to test the validity of our method, we have preformed the same regression described in the text on the M17 sources ($\sim80\%$ of our combined sample) and applied it to the remaining sources acquired from UKIRT.  The C~{\sc iv} velocity offsets were used in the regression since this sample was the largest of the three UV-based redshift estimates.  Prior to correction, the sample of 10 UKIRT sources had a mean, median and $\sigma$ of $-641$ km s$^{-1}$, $-690$ km s$^{-1}$, and $952$ km s$^{-1}$ respectively.  After running the regression on the M17 sample and applying the new correction to the UKIRT sources, the mean, median and $\sigma$ improved to $474$ km s$^{-1}$, $376$ km s$^{-1}$, and $772$ km s$^{-1}$ respectively, demonstrating the validity of our method.\\
\indent The SDSS Pipeline redshift estimate, as noted in P18, is subject to highly uncertain redshift determinations due to lower signal-to-noise ratios or unusual objects.  As seen in our relatively small sample, large redshift discrepancies are apparent particularly in two of the 39 objects that we have with available SDSS Pipeline-based redshifts.  In each case, the velocity offsets are $>10^4$ km s$^{-1}$ and, when included in the regression analysis, it nearly tripled the uncertainty on the redshift determination.  The most robust redshift determination methods involve a correction based on the C~{\sc iv} spectral properties and UV continuum luminosity to either C~{\sc iv}- or HW10-based redshifts.  P18 also provides a redshift based off of visual inspection, $z_{\rm VI}$.  We find that this estimate, where available, provides a much more reliable redshift estimate than the one provided by the SDSS Pipeline.  The mean systematic offset for this redshift estimate is $-290$ km s$^{-1}$ with a standard deviation of $762$ km s$^{-1}$.\\
\indent Regarding the two sources with extremely large velocity offsets, SDSS J142243.02$+$441721.2 and \hbox{SDSS J115954.33$+$201921.1}, we note that our best corrections for their UV-based redshifts provide only modest improvements to the redshift determinations, and that their negative velocity offsets (i.e., blueshifts) take on positive velocity offsets (i.e., redshifts), after the correction is applied.  The velocity offsets for SDSS J142243.02$+$441721.2 improve from \hbox{$-5097$ km s$^{-1}$} to \hbox{$2300$ km s$^{-1}$}, \hbox{$-7740$ km s$^{-1}$} to \hbox{$6016$ km s$^{-1}$}, and \hbox{$-16384$ km s$^{-1}$} to \hbox{$11848$ km s$^{-1}$} for C~{\sc iv}-, HW10-, and SDSS Pipeline-based redshift estimates, respectively.  Similarly, the velocity offsets for \hbox{SDSS J115954.33$+$201921.1} changed from \hbox{$-1264$ km s$^{-1}$} to \hbox{$-58$ km s$^{-1}$}, \hbox{$407$ km s$^{-1}$} to \hbox{$-656$ km s$^{-1}$}, and \hbox{$-10642$ km s$^{-1}$} to \hbox{$8720$ km s$^{-1}$}, respectively.  While most of the corrected velocity offsets are closer to zero, they do not improve appreciably and still affect the statistics significantly.\\
\indent The origin for the abnormally large velocity offset of the SDSS Pipeline redshift of SDSS J115954.33$+$201921.1 most likely stems from the misidentification of the emission lines in the SDSS spectra by the SDSS Pipeline, as discussed in \ref{sec:out}.  As for SDSS J142243.02$+$441721.2, the origin of the large velocity offset of the C~{\sc iv}-based redshift is intrinsic to the quasar and this should not be confused with the coincidental abnormally large velocity offset stemming from the failure of the SDSS Pipeline to correctly identify the UV spectral features (see Sec. 3.1).  Our measured velocity offset of the C~{\sc iv} line ($-5097$ km s$^{-1}$) is consistent, within the errors, with the value reported in Table 6 of \citet{2018A&A...617A..81V} for the source ($-4670$ km s$^{-1}$).  Such sources may point to additional spectral parameters that should be taken into account in future prescriptions for UV-based redshift corrections.  While such objects may be rare ($\lesssim 5$\% in our combined sample), their potential effects on future redshift estimates should be scrutinized to ensure that redshift corrections for the general quasar population are not skewed.  The difficulty in correcting the UV-based redshift of SDSS J142243.02$+$441721.2 is also manifested by the HW10-based redshift which is unable to improve the estimate but rather provides a larger velocity offset (\hbox{$-7740$ km s$^{-1}$}) with respect to the C~{\sc iv}-based value ($-5097$ km s$^{-1}$).\\
\indent With our combined sample of 55 high-redshift quasars, we verify large velocity offsets between UV-based redshift estimates and $z_{\rm sys}$.  Our calibrations to the UV-based redshift estimates can be used to establish more reliable estimates for $z_{\rm sys}$ when working with high-redshift quasars in the optical band.  This effort will lead to more reliable constraints on a range of measurements that require precise distances for quasars.\\
\newpage
\section{Conclusions}\label{sec:5}
In the coming decade, $\approx10^6$ high-redshift ($z\gtrsim 0.8$) quasars will have their redshifts determined through large spectroscopic surveys conducted in the visible band (i.e., rest-frame UV band), e.g., the DESI survey \citep[e.g.,][]{2013arXiv1308.0847L,2016arXiv161100036D}. Many of these quasars, at $1.5 \lesssim z \lesssim 6.0$, will have the prominent C~{\sc iv} emission line covered in their spectra. The spectral properties of this line can provide a valuable means for correcting UV-based redshifts as we have shown in this work.\\
\begin{deluxetable}{lcccc}
\tablecaption{Correction Statistics \label{tab:stats}}
\tablewidth{0pt}
\tablehead{
\colhead{Model} & \colhead{Mean}  & \colhead{Median} & \colhead{$\sigma$} & \colhead{Skew} 
}
\startdata
 C~{\sc iv}\tablenotemark{a} & -1016 & -1028 & 1132 (993) & -1.11\\
 C~{\sc iv} 1 & -20 & -194 & 885 (792) & 0.55 \\
 C~{\sc iv} 2 & -3 & 18 & 837 (755) & 0.66 \\
 C~{\sc iv} 3 & 1 & -80 & 1022 (905) & 0.67\\
 \textbf{C~{\sc iv} 4} & \textbf{0} & \textbf{-24} & \textbf{750\tablenotemark{b} (679)} & \textbf{0.37}\\
 \hline
  HW10\tablenotemark{c} & -121 & 159 & 1310 (719) & -4.09\\
 HW 1 & -14 & -116 & 1123 (575) & 3.86\\
 HW 2 & -2 & -97 & 1157 (638) & 3.38\\
 HW 3 & 1 & -73 & 1195 (621) & 3.98\\
 \textbf{HW 4} & \textbf{1} & \textbf{-68} & \textbf{1067 (547\tablenotemark{b})} & \textbf{3.59}\\
 \hline
 SDSS Pipe\tablenotemark{d} & -1029 & -63 & 3255 (1264) & -3.45 \\
 Pipe 1 & -31 & -558 & 2954 (1161) & 2.78\\
 Pipe 2 & -8 & -578 & 2928 (1165) & 2.66\\
 Pipe 3 & -2 & -697 & 3072 (1200) & 3.03\\
 \textbf{Pipe 4} & \textbf{-3} & \textbf{-449} & \textbf{2851 (1131)} & \textbf{2.54}
 \enddata
 \tablenotetext{a}{55 objects were used in the full correction statistics and 54 objects were used in the correction statistics excluding outliers.}
 \tablenotetext{b}{The best results, with and without outliers, are further discussed in Section \ref{sec:4}.}
  \tablenotetext{c}{50 objects were used in the full correction statistics and 49 objects were used in the correction statistics excluding outliers.}
   \tablenotetext{d}{39 objects were used in the full correction statistics and 37 objects were used in the correction statistics excluding outliers.}
  \tablecomments{Bold results are presented in Figure \ref{fig:voff}.  The $\sigma$ reported in parenthesis is the standard deviation once outliers have been removed.  For C~{\sc iv} and HW10, only SDSS J142243.02$+$441721.2 was removed.  For SDSS Pipe, SDSS J142243.02$+$441721.2 and \hbox{SDSS J115954.33$+$201921.1 were removed}.}
\end{deluxetable}
\indent Using a sample of 55 quasars, our prescription for correcting UV-based redshifts yields a mean systematic velocity offset which is consistent with zero and further improves the uncertainty on the redshift determination by $\sim25 - 35$\% with respect to the method of HW10.  We also find that UV-based redshifts derived from the SDSS Pipeline provide the least reliable results, and the associated uncertainties with respect to $z_{\rm sys}$ cannot be reduced appreciably.  With a larger, uniform sample of high-redshift quasars with NIR spectroscopy (e.g., Matthews et al., in prep.), we plan to improve the reliability of our redshift estimates further and search for additional spectral properties that may further improve these estimates.\\
\indent We show that the uncertainties on UV-based redshifts for the majority of high-redshift quasars can be reduced considerably by obtaining NIR spectroscopy of a larger sample of sources and using the [O~{\sc iii}]-based systemic redshift to inform a C~{\sc iv}-based regression analysis. The reduction in redshift uncertainties is particularly useful for a range of applications involving accurate cosmological distances.
\section{Acknowledgments}
We gratefully acknowledge the financial support by National Science Foundation grants AST-1815281 (C.~D., O.~S.), and AST-1815645 (M.~S.~B., A.~D.~M.).  A.D.M.\ was supported by the Director, Office of Science, Office of High Energy Physics of the U.S. Department of Energy under Contract No.\ DE-AC02-05CH1123 and Award No.\ DE-SC0019022.  This research has made use of the NASA/IPAC Extragalactic Database (NED), which is operated by the Jet Propulsion Laboratory, California Institute of Technology, under contract with the National Aeronautics and Space Administration, as well as NASA's Astrophysics Data System Bibliographic Services.


\newpage
\begin{thebibliography}{}

\bibitem[Bessell et al.(1998)]{1998A&A...333..231B} Bessell, M.~S., Castelli, F., \& Plez, B.\ 1998, \aap, 333, 231

\bibitem[Bolton et al.(2012)]{2012AJ....144..144B} Bolton, A.~S., Schlegel, D.~J., Aubourg, {\'E}., et al.\ 2012, \aj, 144, 144

\bibitem[Boroson(2005)]{2005AJ....130..381B} Boroson, T.\ 2005, \aj, 130, 381.

\bibitem[Boroson \& Green(1992)]{1992ApJS...80..109B} Boroson, T.~A., \& Green, R.~F.\ 1992, \apjs, 80, 109

\bibitem[Chen et al.(2014)]{2014ApJS..215...12C} Chen, Z.-F., Qin, Y.-P., Qin, M., et al.\ 2014, The Astrophysical Journal Supplement Series, 215, 12.

\bibitem[DESI Collaboration et al.(2016)]{2016arXiv161100036D} DESI Collaboration, Aghamousa, A., Aguilar, J., et al.\ 2016, arXiv e-prints, arXiv:1611.00036

\bibitem[Font-Ribera et al.(2013)]{2013JCAP...05..018F} Font-Ribera, A., Arnau, E., Miralda-Escud{\'e}, J., et al.\ 2013, \jcap, 5, 018

\bibitem[Gaskell(1982)]{1982ApJ...263...79G} Gaskell, C.~M.\ 1982, \apj, 263, 79

\bibitem[Gibson et al.(2009)]{2009ApJ...692..758G} Gibson, R.~R., Jiang, L., Brandt, W.~N., et al.\ 2009, \apj, 692, 758

\bibitem[Hewett, \& Wild(2010)]{2010MNRAS.405.2302H} Hewett, P.~C., \& Wild, V.\ 2010, \mnras, 405, 2302.

\bibitem[Hogg(1999)]{1999astro.ph..5116H} Hogg, D.~W.\ 1999, arXiv e-prints, astro-ph/9905116

\bibitem[Hopkins \& Elvis(2010)]{2010MNRAS.401....7H} Hopkins, P.~F., \& Elvis, M.\ 2010, \mnras, 401, 7

\bibitem[Hutchings et al.(2006)]{2006AJ....131..680H} Hutchings, J.~B., Cherniawsky, A., Cutri, R.~M., et al.\ 2006, \aj, 131, 680.

\bibitem[Kaspi et al.(2007)]{2007ApJ...659..997K} Kaspi, S., Brandt, W.~N., Maoz, D., et al.\ 2007, \apj, 659, 997

\bibitem[Levi et al.(2013)]{2013arXiv1308.0847L} Levi, M., Bebek, C., Beers, T., et al.\ 2013, arXiv:1308.0847

\bibitem[Mason et al.(2017)]{2017MNRAS.469.4675M} Mason, M., Brotherton, M.~S., \& Myers, A.\ 2017, \mnras, 469, 4675.

\bibitem[Ofek et al.(2006)]{2006ApJ...641...70O} Ofek, E.~O., Maoz, D., Rix, H.-W., et al.\ 2006, The Astrophysical Journal, 641, 70

\bibitem[P{\^a}ris et al.(2014)]{2014A&A...563A..54P} P{\^a}ris, I., Petitjean, P., Aubourg, {\'E}., et al.\ 2014, \aap, 563, A54

\bibitem[P{\^a}ris et al.(2017)]{2017A&A...597A..79P} P{\^a}ris, I., Petitjean, P., Ross, N.~P., et al.\ 2017, \aap, 597, A79

\bibitem[P{\^a}ris et al.(2018)]{2018A&A...613A..51P} P{\^a}ris, I., Petitjean, P., Aubourg, {\'E}., et al.\ 2018, \aap, 613, A51

\bibitem[Prochaska et al.(2013)]{2013ApJ...776..136P} Prochaska, J.~X., Hennawi, J.~F., Lee, K.-G., et al.\ 2013, \apj, 776, 136 

\bibitem[Richards et al.(2009)]{2009ApJS..180...67R} Richards, G.~T., Myers, A.~D., Gray, A.~G., et al.\ 2009, The Astrophysical Journal Supplement Series, 180, 67.

\bibitem[Shemmer \& Lieber(2015)]{2015ApJ...805..124S} Shemmer, O., \& Lieber, S.\ 2015, \apj, 805, 124

\bibitem[Shen et al.(2011)]{2011ApJS..194...45S} Shen, Y., Richards, G.~T., Strauss, M.~A., et al.\ 2011, \apjs, 194, 45.

\bibitem[Shen et al.(2016)]{2016ApJ...831....7S} Shen, Y., Brandt, W.~N., Richards, G.~T., et al.\ 2016, \apj, 831, 7.

\bibitem[Sheskin(2007)]{sheskin07} Sheskin, D. J.\ 2007, Handbook of Parametric and Nonparametric Statistical Procedures, (4th ed.; Chapman \& Hall/CRC)

\bibitem[Schneider et al.(2010)]{2010AJ....139.2360S} Schneider, D.~P., Richards, G.~T., Hall, P.~B., et al.\ 2010, \aj, 139, 2360

\bibitem[Skrutskie et al.(2006)]{2006AJ....131.1163S} Skrutskie, M.~F., Cutri, R.~M., Stiening, R., et al.\ 2006, \aj, 131, 1163.

\bibitem[Spergel et al.(2007)]{2007ApJS..170..377S} Spergel, D.~N., Bean, R., Dor{\'e}, O., et al.\ 2007, \apjs, 170, 377.

\bibitem[Tytler \& Fan(1992)]{1992ApJS...79....1T} Tytler, D., \& Fan, X.-M.\ 1992, \apjs, 79, 1 

\bibitem[Vanden Berk et al.(2001)]{2001AJ....122..549V} Vanden Berk, D.~E., Richards, G.~T., Bauer, A., et al.\ 2001, \aj, 122, 549.

\bibitem[Vietri et al.(2018)]{2018A&A...617A..81V} Vietri, G., Piconcelli, E., Bischetti, M., et al.\ 2018, \aap, 617, A81

\bibitem[York et al.(2000)]{2000AJ....120.1579Y} York, D.~G., Adelman, J., Anderson, J.~E., Jr., et al.\ 2000, \aj, 120, 1579 

\bibitem[Zhao et al.(2019)]{2019MNRAS.482.3497Z} Zhao, G.-B., Wang, Y., Saito, S., et al.\ 2019, \mnras, 482, 3497

\end{thebibliography}
\end{document}